\documentclass[12pt]{article}

\usepackage[english]{babel}
\usepackage{amsmath}
\usepackage{amsthm}
\usepackage{amssymb}
\usepackage{enumerate}
\usepackage{xspace}
\usepackage{euscript}
\usepackage{graphicx}    
\usepackage{amscd}
\usepackage{epsfig}
\usepackage{tabularx}
 \usepackage{color}
\usepackage[all]{xy}

\begin{document}
\title{\bf Scattering of a two skyrmion configuration  on potential holes or barriers in a model Landau-Lifshitz equation}
\author{J.C. Collins\footnote{J.C.Collins@durham.ac.uk} and  W.J. Zakrzewski\footnote{W.J.Zakrzewski@durham.ac.uk} 
\\ Department of Mathematical Sciences, University of Durham, \\ Durham DH1 3LE, UK}
\date{}
\maketitle

\begin{abstract}

The dynamics of a baby skyrmion configuration, in a \\model Landau-Lifshitz equation, was studied in the presence of  various potential obstructions. The baby skyrmion configuration was constructed from two $Q=1$ hedgehog solutions to the baby skyrme model in (2+1)dimensions. The potential obstructions were created by introducing a new term into the Lagrangian which resulted in a localised inhomogeneity in the potential terms' coefficient. In the barrier system, the normal circular path was deformed as the skyrmions traversed the barrier. During the same period, it was seen that the skyrmions sped up as they went over the barrier. For critical values of the barrier height and width, the skyrmions were no longer bound and were free to separate. In the case of a potential hole, the baby skyrmions no longer formed a bound state and moved asymptotically along the axis of the hole. It is shown how to modify 
the definition of the angular momentum to include the effects of the obstructions, so that it is conserved.

\end{abstract}


\renewcommand{\thesection}{\Roman{section}}

\section{Introduction}

The scattering of particles off potential holes and barriers in classical and quantum mechanical systems are seemingly different. In a classical system, if the particle has sufficient energy it can traverse a barrier; if it does not, it gets reflected. When it encounters a hole it speeds up as it passes over the hole and is always transmitted. In both cases the particle is either transmitted or reflected. In quantum mechanical scattering the particle can be reflected and transmitted for either a barrier or a hole but these events occur with a certain probability which is dependent on the particle's energy and on the size of the barrier or hole.

In this paper we examine the scattering properties  of a topological soliton in magnetic systems, whose motion is then governed by the Landau-Lifshitz equation. Topological solitons are, of course, classical objects (as they satisfy classical equations of motions). However, they describe extended objects
and, as shown in \cite{Wojtek4}, some of their properties resemble those of quantum systems. Hence in this paper we study 
this problem further; this time concentrating on systems whose dynamics is described by a Landau Lifshitz equation.
This equation arises in the dynamics of magnetic bubbles and so, in this paper, we look at the behaviour of topological solitons
in the presence of potential obstructions.
The topological solitons under investigation are baby skyrmions which are thought to describe the experimentally observable magnetic bubbles. 

Our investigation should also shed some light on the properties of magnetic bubbles. Such bubbles are not mathematical artifacts but
have been produced experimentally by subjecting a ferromagnetic material to a pulsed magnetic field. The ferromagnetic domains of the material are then squeezed by the field. If a field of large enough magnitude is applied over a sufficient length of time the magnetic domains of the system collapse leaving behind the material which is uniformly magnetised in the direction of the applied field. This process is not instantaneous or uniform. As the domains tend to align with the applied field, the pulsing of certain materials results in the pinching of the domain walls into a cylindrical domain called a magnetic bubble. Magnetic bubbles were once considered as an alternative form of memory storage due to the density at which they could be stored.

 In real three dimensional systems the magnetic bubbles are stabilised by the finite thickness of the thin films in which they are created. The model examined here is strictly in 2-D. In this case the bubbles are stabilised by the introduction of the slightly artificial Skyrme term. For more realistic calculations one would need to perform 3-D simulations. Magnetic bubbles and their properties have been extensively experimentally researched; further information on this research can be found in \cite{bobeck} and \cite{Odell}.

The baby skyrme model in (2+1)-dimensions is defined by:
\begin{eqnarray}
\mathcal{L} = \frac{1}{2}\gamma_{1}\partial_{\mu}{\underline{\phi}}\cdot \partial^{\mu}\underline{\phi}- \frac{1}{4}\gamma_{2}[({\partial_{\mu}{\underline{\phi}}\cdot\partial^{\mu}\underline{\phi}})^{2}-(\partial_{\mu}\underline{\phi}\cdot\partial_{\nu}\underline{\phi})(\partial^{\mu}\underline{\phi}\cdot\partial^{\nu}\underline{\phi})]- V(\underline \phi) , \label{L} 
\end{eqnarray}
where $\underline{\phi}$ is a 3-component scalar field and the indices run over the space-time coordinates. This is referred to as the baby skyrme model to distinguish it from the full skyrme nuclear model of baryons. The first term is the exchange energy, the second and third terms are the skyrme term and the potential term respectively. The latter two terms have been introduced to avoid the consequences of Derrick's theorem \cite{derrick} and to stabilise topological soliton solutions in 2-dimensions. The condition that $\underline\phi^{2}=1$ is imposed so that the target space is the 2-sphere, such that $\underline{\phi}$ is now a map $\underline{\phi}: \mathbb {R}^{2} \rightarrow {S}^{2}$. For finite energy solutions it is necessary for the fields to tend to a vacuum  at infinity, where $\phi_{3}=1$ at $\infty$. This results in a compactification of $ \mathbb{R}^{2}$ so that $\underline{\phi}$ now takes values in the extended plane $ \mathbb{R}^{2}\cup \infty $, which is topologically equivalent to  $S^{2}$. The constraint equation $\underline\phi^{2}=1$ and the boundary condition at infinity results in the field $\underline \phi$ becoming a non-trivial map $\underline{\phi}:S^{2}\rightarrow S^{2}$. Each soliton solution is grouped into a different homotopy class according to the winding number, or topological charge, of this map. The topological charge Q is given by:
\begin{equation}
{Q}=\frac{1}{8 \pi}\int_{-\infty}^{\infty}\epsilon_{ij} \underline{\phi}\cdot(\partial_{j}\underline{\phi} \times \partial_{i}\underline{\phi})d^{2}x , \label{top}
\end{equation}
 where the indices $i,j$ run over the space coordinates and $Q \in \mathbb{Z}$. The topological soliton solutions of the baby skyrme model are called baby skyrmions. Here, for simplicity, we shall refer to baby skyrmions of charge Q as Q-skyrmions.
Recent work has shown that the continuum dynamical equation in anti-ferromagnetic systems, also resembles a second order relativistic wave equation \cite{speight} in which such solitons solutions exist. The work in \cite{speight} established many interesting properties of solitons in such systems.

Our discussion so far has concerned the relativistic systems of skyrmions. However, as we said above, they
also arise in the description of magnetic bubbles, but this time their evolution is described by
the first order Landau-Lifshitz equation.  The Landau-Lifshitz equation is given by:

\begin{equation}
\frac{\partial\underline{\phi}}{\partial{t}}= \underline{\phi} \times\frac{\delta{W}}{\delta\underline{\phi}},  \label{LL}
\end{equation}
 where $W$ is the energy functional written as:
\begin{equation}
W =\iint_{- \infty}^{\infty} w \,dx \,dy , \label{W}
\end{equation}
and $w$ is the static part of (\ref{L}) given by:
\begin{equation}
w =\frac{1}{2}\gamma_{1}\partial_{i}{\underline{\phi}}\cdot \partial_{i}\underline{\phi}+ \frac{1}{4}\gamma_{2}[({\partial_{i}{\underline{\phi}}\cdot\partial_{i}\underline{\phi}})^{2}-(\partial_{i}\underline{\phi}\cdot\partial_{j}\underline{\phi})(\partial_{i}\underline{\phi}\cdot\partial_{j}\underline{\phi})] + V(\underline{\phi}). \nonumber
\end{equation}

Thus we can write down $\frac{\delta{W}}{\delta\underline{\phi}}$:

\begin{eqnarray}
\frac{\delta{W}}{\delta\underline{\phi}}&=& \gamma_{1}\nabla^{2}\underline{\phi} - \gamma_{3}\frac{\partial V(\underline{\phi})}{\partial \underline{\phi}} \nonumber \\
& +  & \frac{1}{2}\gamma_{2}\big\{2\partial_{i}\big[(\partial_{j}\underline{\phi}\cdot\partial_{j}\underline{\phi})\partial_{i}\underline{\phi} \big]
- \partial_{i}\big[(\partial_{i}\underline{\phi}\cdot\partial_{j}\underline{\phi})\partial_{j}\underline{\phi} \big]
- \partial_{j}\big[(\partial_{i}\underline{\phi}\cdot\partial_{j}\underline{\phi})\partial_{i}\underline{\phi} \big] \big\}. \nonumber
\end{eqnarray}

Analysis of the dynamics in Landau-Lifshitz systems has been greatly simplified by the work of Papanicolaou and Tomaras \cite{papatom}, who constructed unambiguous conservation laws for the system governed by (\ref{LL}). In their work they found that the important quantity was the topological charge density $q$:
\begin{equation}
{q}=\epsilon_{ij} \underline{\phi}\cdot(\partial_{j}\underline{\phi} \times \partial_{i}\underline{\phi}). \label{topdens}
\end{equation}
The conservation laws are constructed from the moments of q. They involve:
\begin{eqnarray}
l&=&\frac{1}{2} \iint_{- \infty}^{\infty} \underline{x}^{2}q \,dx  \,dy ,\label{l}\\
m&=&\iint_{- \infty}^{\infty} (\phi_{3}-1)  \,dx  \,dy \label{m}, \\
J&=&l+m ,
\end{eqnarray}
where $l$ is the orbital angular momentum, $m$ is the total magnetization in the third direction and $J$ is the total angular momentum.
 Conservation laws for the system were constructed by examining the time evolution of q:
\begin{equation}
{\dot {q}}=-\epsilon_{ij}\partial_{i}\partial_{l}\sigma_{jl} , \label{qdot}
\end{equation}
where $\partial_{l}\sigma_{jl}$ can be written in terms of the energy functional $W$:
\begin{equation}
\partial_{l}\sigma_{jl}=\left ( \frac{\delta W}{\delta \underline{ \phi}}\cdot{\partial_{j} \underline \phi}\right ) ,
\end{equation}
Taking an explicit time derivative of (\ref{l}) gives:
 \begin{equation}
\dot{l}=\frac{1}{2} \iint_{- \infty}^{\infty}\underline{x}^{2}\dot {q} \,dx  \,dy ,\label {ldot}
\end{equation}
and we note that this can be recast as:
 \begin{equation}
\dot{l}= \iint_{- \infty}^{\infty} \epsilon_{ij}\sigma_{ij}  \,dx  \,dy ,\label {l dot tensor}
\end{equation}
by integrating (\ref{ldot}) by parts. In the case of a system with a symmetric tensor  $\sigma_{ij}$ , $\dot{l}=0$ and angular momentum is conserved.

 The guiding centre coordinate $\underline{R}$ of the soliton is defined as the first moment of the topological charge density $q$:
\begin{equation}
\underline{R}= \frac{1}{4 \pi Q} \iint_{- \infty}^{\infty}\underline{x} q \,dx  \,dy . \label {R}
\end{equation}
Since our solitons are in two spatial dimensions, their position can generally be defined as the location of the centre of each soliton. This definition can be interpreted in two ways. One can consider the soliton centre to be the point at which the third component of the field $\phi_{3}= -1$ or to be the maxima of the topological charge density $q$. It was seen in previous simulations \cite{Wojtek2} that both of these definitions produce near identical trajectories. One can also consider the mean squared radius $r$ of the solitons, defined by:
\begin{equation}
{r}^{2}= \frac{1}{4 \pi Q} \iint_{- \infty}^{\infty}\big( \underline{x}-\underline{R}\big)^{2}q \,dx  \,dy .\label {R}
\end{equation}
One can expand out (\ref{R}) to find a relationship between the mean squared radius of the solitons and $l$:
\begin{equation}
r^{2}= \frac{l}{2 \pi Q}- \underline{R}^{2}. \label{r1}
\end{equation}

This relationship between $l$, $r$ and $R$ greatly helps to understands the dynamics described in later sections.
\\
Much of the previous work on baby skyrme models, in the context of Landau-Lifshitz dynamics, has concerned the choice of two different potential terms $V(\phi)$:
\begin{eqnarray}
V(\phi)&=&\frac{1}{2} \gamma_{3}(1-\phi_{3})^{4}, \label{old baby}\\
V(\phi)&=&\frac{1}{2} \gamma_{3}(1-{\phi_{3}}^{2}). \label{new baby}
\end{eqnarray}

Models which employ either of these potentials are commonly referred to as the holomorphic baby skyrme model (\ref{old baby}) and the `new' baby skyrme model (\ref{new baby}) to distinguish it from the holomorphic one. The holomorphic model was first studied in the context of Landau-Lifshitz dynamics in \cite{Wojtek2}, since it provided an analytical solution to the system of equations. The topological solitons of this model are polynomially localised. In \cite{Wojtek2} it was shown that two 1-skyrmions orbited around each other along deformed circular trajectories. Their work involved a local magneto-static field in addition to the three previous terms in (\ref{L}), which resulted in the non-conservation of $l$ and $m$. The total angular momentum $J$ was well conserved in time. The authors of \cite{Wojtek2} attributed the non-conservation of $l$ and $m$ to the non-symmetric structure of $\sigma_{jl}$ due to the presence of the magneto-static field.

The new baby skyrme model is a more realistic case of easy axis anisotropy and we use this in our study i.e. we consider $V(\phi)$ to take the form of (\ref{new baby}). Some of the work done on it has been in relation to the dynamics of magnetic bubbles \cite{Wojtek1}; this study also involved a local magneto-static field. Analytic solutions do not exist and so all solutions must be found numerically. The topological solitons of this model are exponentially localised. In \cite{Wojtek1} it was found that two $Q=1$-skyrmions orbited each other on a circular trajectory modified by a Larmor procession due to the magnetic field. $J$ was well conserved in time but its constituent components $l$ and $m$ were not. The arguments for the non-conservation were identical to those in the holomorphic model. Since those early papers, there has been extensive work done on both these models and their multiskyrmion structures. The details can be found in the work of Weidig \cite{Weidig}, or for a larger class of potentials $V(\underline{\phi})$ in \cite{Wojtek3}.


\section{Constructing the initial field configuration}

Seeking a static field configuration which is a solution of (\ref{LL}) for a potential term $V(\phi)$ of the form (\ref{new baby}), we assume the solitons take the form of a hedgehog configuration for a Q-skyrmion:
\begin{equation}
\phi = (\cos(Q\theta)\sin(f(r)),\sin(Q\theta)\sin(f(r)),\cos(f(r))), \label{hog}
\end{equation}
 where $f(r)$ is the profile function satisfying certain boundary conditions, $\theta$ is the polar angle and $Q$ is the topological charge of the skyrmions. The skyrmion solutions are minima of the energy functional (\ref{W}). We are interested in the dynamics of $Q=1$ skyrmions.  Inserting the hedgehog configuration for $Q=1$ into the energy functional and minimising the integral, results in a second order differential equation for the profile function $f(r)$:
\begin{eqnarray}
&{f^{\prime\prime}}&\left(\gamma_{1}r+\frac{\gamma_{2}\sin^{2}f}{r}\right)+{f^{\prime}}\left(\gamma_{1}-\frac{\gamma_{2}\sin^{2}f}{r^{2}}\right) \nonumber\\
&+&{f^{\prime}}^{2}\left(\frac{\gamma_{2}\sin f\cos f}{r}\right) - \frac{\gamma_{1}\sin f\cos f}{r}- \gamma_{3}r \cos{f} \sin{f}=0 .\label{ODE}
\end{eqnarray}
This can be rearranged into the form $f^{\prime\prime}=h(r,f,f^\prime)$. The profile function must satisfy certain boundary conditions for there to exist finite energy solutions. The boundary conditions impose constraints on $f(r)$ at  the origin and infinity: $f(0)=\pi$ and $f(\infty)=0$. The second order differential equation for $f(r)$ can be solved numerically using the shooting method. With the profile function obtained we can construct from (\ref{hog}) a 1-skyrmion solution to (\ref{LL}). A two 1-skyrmion configuration can be constructed by the superposition procedure. The easiest method to do this is to transform the fields to a stereographic variable $\Omega$:
\begin{eqnarray}
\Omega &=& \frac{\phi_{1}+i\phi_{2}}{1+\phi_{3}} , \nonumber  \\
\phi_{1}&=&\frac{\Omega+\Omega^{*}}{1+{|\Omega|}^{2}} , \nonumber\\
\phi_{2}&=&\frac{1}{i}\frac{\Omega-\Omega^{*}}{1+{|\Omega|}^{2}} , \label{stereo} \\
\phi_{3}&=&\frac{1-|\Omega|^{2}}{1+|\Omega|^{2}} . \nonumber
\end{eqnarray}
 The stereographic variable $\Omega$ can then be rewritten in terms of the hedgehog anzatz variables, $f(r)$ and $\theta$ as: 
\begin{equation}
\Omega=\tan(\frac{f({r})}{2})e^{i\theta} . \nonumber
\end{equation}
To construct a two $Q=1$-skyrmion configuration in which the two skyrmions are in an attractive channel, we take:
\begin{eqnarray}
\Omega &=& \Omega_{1} - \Omega_{2} , \label{config}\\
\Omega_{1}&=&\tan(\frac{f(r_{1})}{2})e^{i\theta_{1}} , \nonumber\\
\Omega_{2}&=&\tan(\frac{f(r_{2})}{2})e^{i\theta_{2}} , \nonumber
\end{eqnarray}
where $r_{i}=\sqrt{\big( (x-x_{i})^{2}+(y-y_{i})^{2} \big)}$ and $\theta_{i}=Tan^{-1}\big(\frac{y-y_{i}}{x-x_{i}}\big)$ are calculated relative to the centres of the skyrmions $(x_{i},y_{i})$. During the simulations it was found that the configuration constructed in this manner did not replicate the skyrmion ring configurations for small values of \\ $d=\sqrt{\big( (x_{1}-x_{2})^{2}+(y_{1}-y_{2})^{2} \big)}$; see \cite{Weidig}. The superposition procedure was a very good approximation to a two 1-skyrmion configuration for values of $d>6$, where the skyrmions were well separated to be distinct. To obtain a true representation of the configuration for all values of d, we used a gradient flow method to `relax' the field configuration. The above field configuration (\ref{config}) constructed for skyrmion separation $d=8$, was used as an initial condition of the gradient flow equation given by:
\begin{equation}
\frac{\partial \underline{\phi}}{\partial t}=-\kappa \frac{\delta W}{\delta \underline{\phi}}+k\phi ,\nonumber
\end{equation}
where $k$ is a Lagrange multiplier introduced such that the constraint $\underline{\phi}^{2}=1$ is satisfied. The field configurations obtained by this relaxation method show the required ring like properties for small values of the skyrmion separation.


\section{Potential obstruction}

In this paper we study the scattering properties of a two $Q=1$-skyrmion configuration on a potential obstruction which is localised in a finite region of space. In constructing the obstruction we adopt a similar approach used in the previous work of one of the authors \cite{Wojtek4} and introduce a term into the Lagrangian (\ref{L}) which vanishes in the vacuum state  $\phi_{3}=+1$. The obstructions need to be introduced in this way so that the tails of the solitons are not changed by the obstruction. Therefore, we add the additional potential term $V_{obstruction}(\phi_{3})$ which is identical to the potential in (\ref{L}) and so it effects the Lagrangian of (\ref{L}) by changing the potential coefficient from a constant to one that depends on the space coordinates. The new Lagrangian of the system written in terms of the previous one:
\begin{equation}
\mathcal{L}_{new}= \mathcal{L}_{old} + V_{obstruction}(\phi_{3}), \nonumber
\end{equation}where:
\begin{equation}
V_{obstruction}(\phi)= \frac{1}{2}\Gamma(1-\phi_{3}^{2}).
\end{equation}
$\Gamma$ can be either positive or negative. The sign of $\Gamma$ determines whether the potential obstruction is a hole or a barrier. When $\Gamma > 0$, the obstruction is a barrier. Conversely when $\Gamma < 0$ the obstruction is a hole. The introduction of this term, implies that (\ref{L}) can be rewritten as:
\begin{equation}
\mathcal{L} = \frac{1}{2}\gamma_{1}\partial_{i}{\underline{\phi}}\cdot \partial_{i}\underline{\phi} - \frac{1}{4}\gamma_{2}[({\partial_{i}{\underline{\phi}}\cdot\partial_{i}\underline{\phi}})^{2}-(\partial_{i}\underline{\phi}\cdot\partial_{j}\underline{\phi})(\partial_{i}\underline{\phi}\cdot\partial_{j}\underline{\phi})] -\frac{1}{2}\gamma_{3}(x,y)(1-\phi_{3}^{2}),
\end{equation}
where the potential term coefficient $\gamma_{3}$ is now a function of the coordinates $(x,y)$ and the static part of (\ref{L}) is only considered as imposed by the Landau-Lifshitz equation (\ref{LL}). This inhomogeneity will be localised to a finite region of space. The value of $\gamma_{3}(x,y)$, in this region, will determine whether the obstruction is a hole or a barrier. In this region, $\gamma_{3}(x,y)$, can therefore be summarised as:
\begin{equation}
\gamma_{3}(x,y)_{region}= \nonumber \gamma_{3}+\Gamma,\qquad 
\begin{cases}
\Gamma>0 & \mbox{Barrier,} \\  
\Gamma<0 & \mbox{Hole.}    
\end{cases}
\end{equation}


\section{Numerical procedures and the free system dynamics}
Unfortunately, it is impossible to solve (\ref{LL}) analytically we have therefore had to study this problem numerically. The fields and their derivatives were discretised in the usual manner and were placed on a lattice of $251\times251$ points, with lattice spacing $dx=0.1$. The numerical integration of the 3-coupled differential equations of (\ref{LL}) involved the use of a $4^{th}$ order Runge-Kutta method 
of simulating time evolution with a time step of $dt=0.001$. The various integrals calculated throughout the simulations were performed using a 2-D Simpson's rule. The constraint equation requires that the fields lie on the 2-sphere, $\underline \phi^{2}=1$, and this was imposed at every time step by rescaling each field component so that $\phi_{i}\rightarrow\frac{\phi_{i}}{\sqrt{\underline \phi \cdot \underline \phi}}$.
 
The skyrmions were initially placed at $(0,\pm d/2)$ in the upper and lower planes, where d is the distance between the two skyrmion centres $(x_{i},y_{i})$. The trajectory of each skyrmion was tracked by following the maxima of the topological charge density and interpolating between the lattice points. All of the simulations have been performed for a skyrmion separation of $d=6$. This was found to be the optimum distance, where the skyrmions are separated enough from each other to be distinct but close enough to interact. The coefficients $\gamma_{i}$ have been set to unity in all the simulations unless stated otherwise.


\subsection{No obstruction dynamics}
\begin{figure}[htbp]
\unitlength1cm \hfil
\begin{picture}(6,6)
 \epsfxsize=7cm \put(0,0){\epsffile{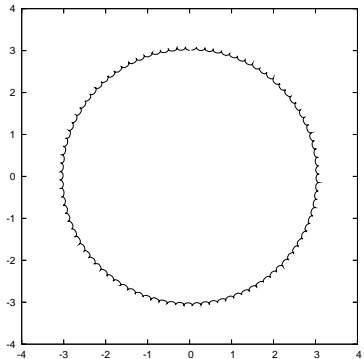}}
\end{picture}
\caption{\label{free}Trajectory of the upper skyrmion in the absence of any obstruction for a two 1-Skyrmion configuration.}
\end{figure}

Initially we examined the behaviour of the skyrmions without an obstruction i.e. with $\Gamma = 0 $. Figure (\ref{free}) shows the trajectory of the upper skyrmion of a two 1-Skyrmion configuration for $\Gamma=0$. The skyrmions orbit around the configuration's centre $(0,0)$. The trajectories of each skyrmion lie along a circle of radius $r\simeq3$. Their position undergoes mild oscillations during the simulation. The total energy $E_{tot}$ and angular momentum $J=l+m$ are conserved with time. Additionally, each individual angular momentum component $l$ and $m$ is also conserved. The time scale for one period is 700 secs. This motion of two baby-skyrmions in a Landau-Lifshitz system is well understood and an analogy with the Hall motion of two interacting electrons is usually invoked when discussing their trajectory \cite{papatom}.


\section{Simple obstruction}
There are many choices one can make for the geometry of the potential obstruction. The simplest choice initially studied was a symmetric obstruction. The obstruction was placed symmetrically along the x-axis with width $b$, i.e. starting at $y=-b/2$  and continuing up to $y=b/2$ and extending for all values of x. In our study we look at the differences of dynamics due to holes and barriers. They will be discussed in the following sections.

\subsection{Potential hole}

\begin{figure}[htbp]
\unitlength1cm \hfil
\begin{picture}(13,13)
 \epsfxsize=10cm \put(3.5,8){\epsffile{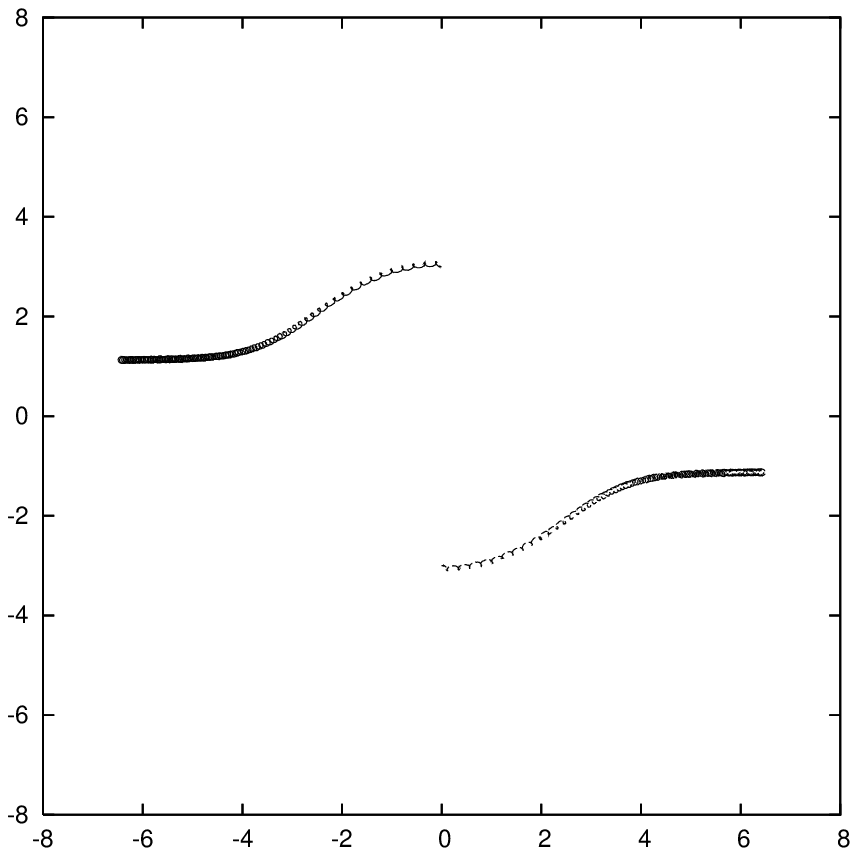}}
 \epsfxsize=10cm \put(0,0.25){\epsffile{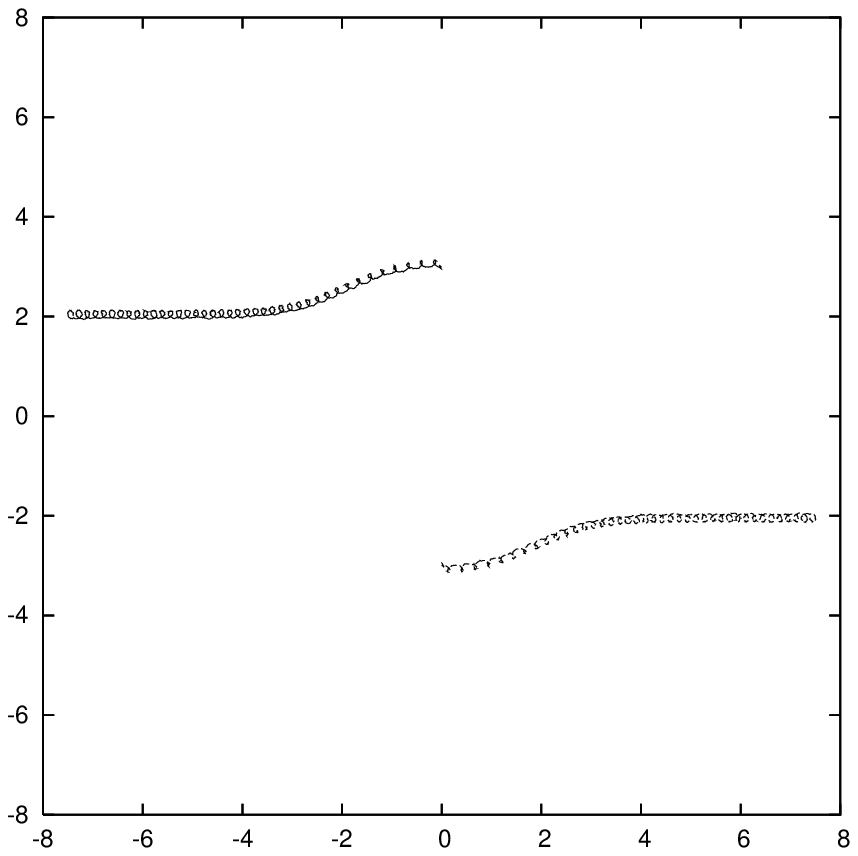}}
 \epsfxsize=10cm \put(7,0.25){\epsffile{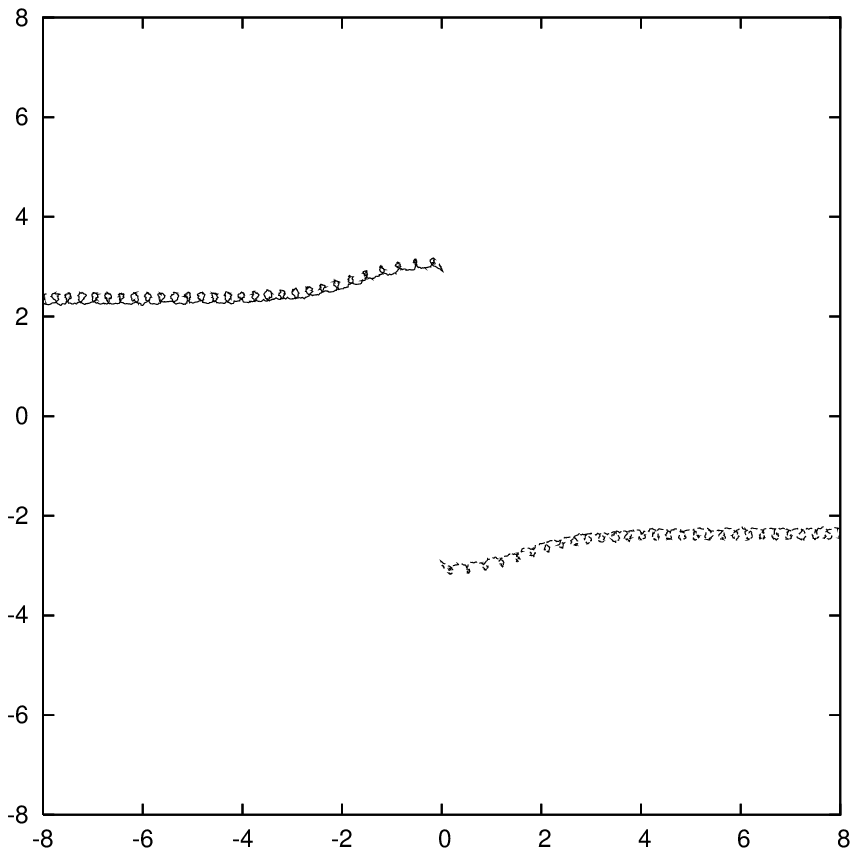}}

\put(7.25,7.5){a}
\put(3.75,0){b}
\put(10.75,0){c}
\end{picture}
\caption{\label{holetraj} Trajectories for both the upper and lower skyrmions with a potential hole, for $b=1$ and with  the time length shown in brackets: a)$\Gamma=-0.1$ (690 secs), b)$\Gamma=-0.25$ (290 secs), c)$\Gamma=-0.5$ (215 secs).}
\end{figure}

\begin{figure}[htbp]
\unitlength1cm \hfil
\begin{picture}(8,8)
\epsfxsize=10cm \put(0,0){\epsffile{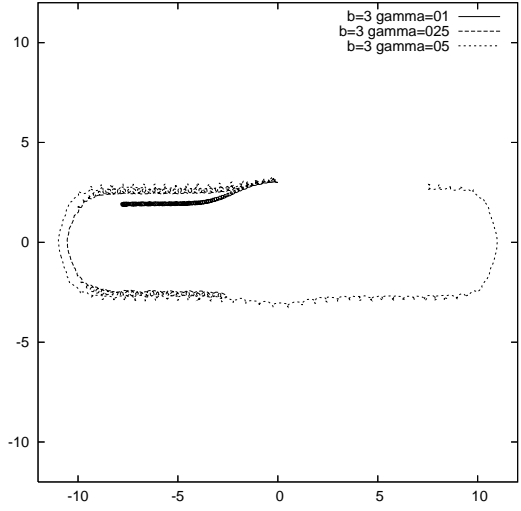}}

\end{picture}
\caption{\label{hole traj} Plot of the trajectories of the upper skyrmion in a system with a potential hole of width b=3 for various values of $\Gamma$.}
\end{figure}

We start by recalling that in the absence of all obstructions the skyrmion execute a circular path around their centre. Fig.(\ref{holetraj}) shows the trajectories of the upper and lower skyrmion encountering a potential hole for different values of $\Gamma$, for $b=1$. In all plots the skyrmions initially try to execute the trajectory of Fig.(\ref{free}) but are deflected. They move asymptotically along the axis of the hole at an approximately  constant value, $y_{max}$. It is clear from Fig.(\ref{holetraj}) that the larger the $|\Gamma|$, the larger the value of $y_{max}$. The skyrmions of Fig.(\ref{holetraj})a are able to get `closer' to the hole than the skyrmions of Fig.(\ref{holetraj})b or Fig.(\ref{holetraj})c. It is reasonable to suggest that the repulsion of the skyrmions by the hole increases with increasing $|\Gamma|$. Fig.(\ref{hole traj}) shows the trajectories of only the upper skyrmions interacting with the potential hole for various values of $\Gamma$ when b=3. Comparing this plot with those in Fig.(\ref{holetraj}) one sees the effect of a larger $b$. The larger the $b$ the larger the $y_{max}$ for a given $\Gamma$. The skyrmions' tail can feel the hole earlier in a system of larger $b$, than in a system of smaller $b$. The same dependence of $y_{max}$ on $\Gamma$ is evident in Fig.(\ref{hole traj}). In Fig.(\ref{hole traj}) the skyrmion trajectories of $\Gamma= -0.25, -0.5$ are seen to be reflected by the boundary. The skyrmions in the lower plane execute similar trajectories.

To explain the observed behavior of skyrmions when encountering a hole, which may at first sight, appear as non-classical, we need to examine the binding energies of the configuration. Table(\ref{1S table}) presents the energies of a single skyrmion in (\ref{LL}), with $\gamma_{3}$ set at the same value throughout the whole lattice. If the energy of the two 1-skyrmion configuration in the presence of a hole is denoted by $E_{2}$ and $E_{1}$ is the energy of a single 1-skyrmion in a system with $\gamma_{3}$ set at the same value as the hole in $E_{2}$, then the binding, or the interaction energy, is given by:
\begin{equation}
 E_{B}=E_{2}-2E_{1} .\label{EB}
\end{equation}
In the system with $b=1$, $\Gamma=-0.1$ we have $E_{2}= 2.1206/8{\pi}$ and $E_{1}= 1.0454/8{\pi}$. The binding energy of the two skyrmions in this system is thus $E_{B}= 0.0356/8{\pi}$, which is positive. Hence in the case corresponding to our simulations the skyrmions are no longer bound. Correspondingly, the system with $b=1$, $\Gamma= -0.5$ has $E_{2}= 2.1205/8{\pi}$ and  $E_{1}= 0.9494/8{\pi}$. The binding energy of the two skyrmions in this system is $E_{B}= 0.2217/8{\pi}$. This is also positive but larger by a factor $\simeq6.3$. Thus both skyrmion configurations placed in the presence of a potential hole are no longer bound. This analysis also explains some of the other features of the system. The trajectories of Fig.(\ref{hole traj})a take substantially more time than Fig.(\ref{hole traj})b or Fig.(\ref{hole traj})c. $E_{B}$ is very much larger when $\Gamma= -0.5$ than for $\Gamma= -0.1$ and, hence, this extra available energy allows the skyrmions in $\Gamma= -0.5$ to separate more quickly. In consequence, the skyrmions for the systems of larger $|\Gamma|$ have a larger speed in this `asymptotic' state than those of smaller $|\Gamma|$.


\begin{table}
\begin{center}

\begin{tabular}{|c|c|}

\hline
$\gamma_{3}$ & E/8$\pi$  \\
\hline \hline
1.5 & 1.0708 \\
1.4 & 1.0705 \\
1.3 & 1.0702 \\
1.2 & 1.0700 \\
1.1 & 1.0697 \\
 1  & 1.0694 \\
0.9 & 1.0454 \\
0.8 & 1.0214 \\
0.7 & 0.9974 \\
0.6 & 0.9734 \\
0.5 & 0.9494 \\

\hline

\end{tabular}

\caption{\label{1S table} Table showing the variation in total energy, in units of $8\pi$, for a 1-skyrmion with the $\gamma_{3}$ coefficient set at the corresponding value all over the whole lattice.}
\end{center}
\end{table}

\newpage


\subsection{Potential barrier}
\begin{figure}[htbp]
\unitlength1cm \hfil
\begin{picture}(13,13)
 \epsfxsize=10cm \put(3.5,8){\epsffile{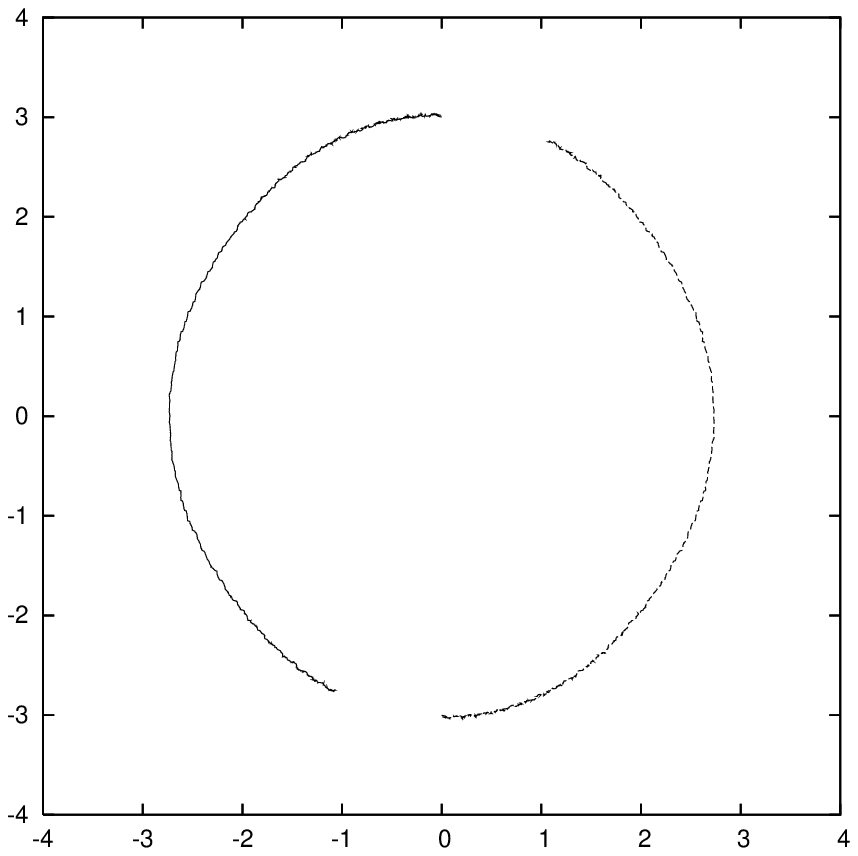}}
 \epsfxsize=10cm \put(0,0.25){\epsffile{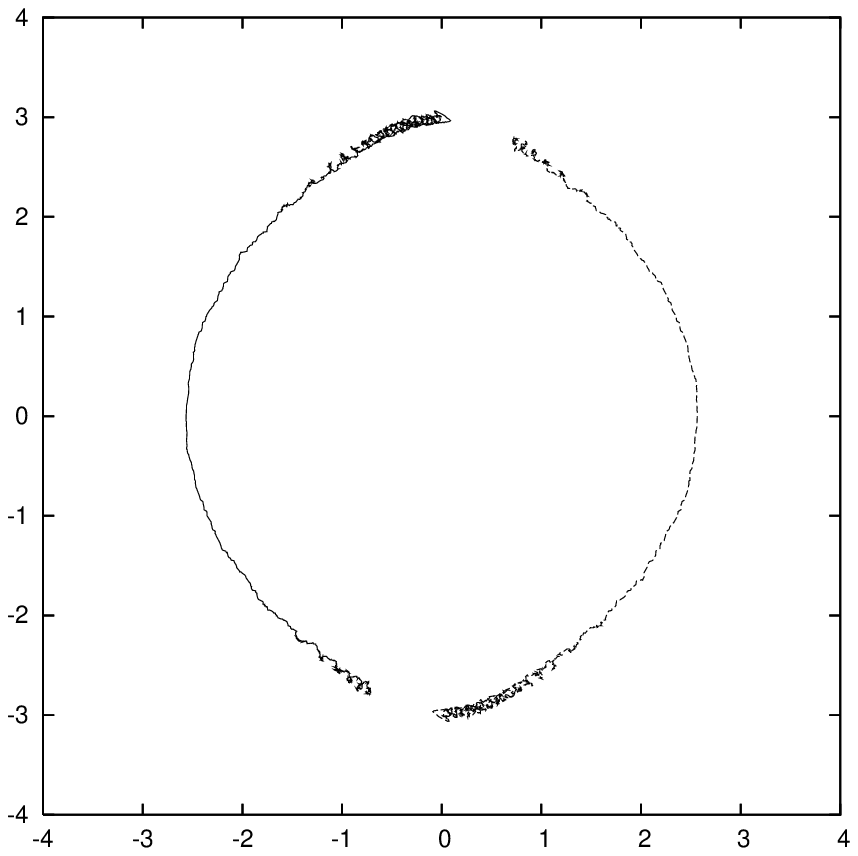}}
 \epsfxsize=10cm \put(7,0.25){\epsffile{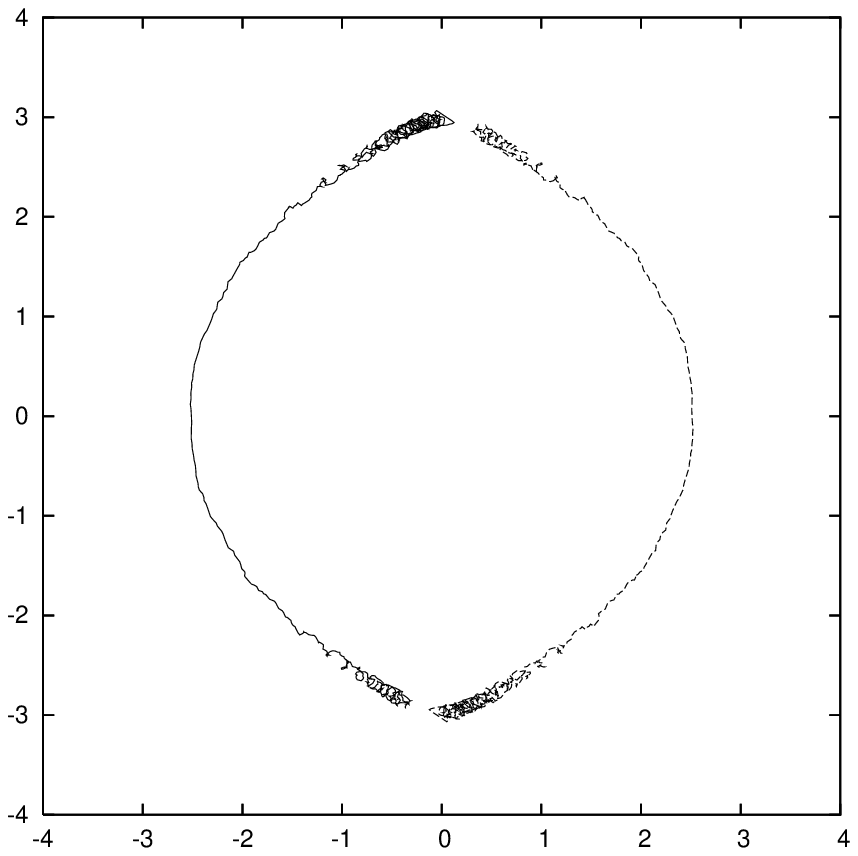}}

\put(7.25,7.5){a}
\put(3.75,0){b}
\put(10.75,0){c}
\end{picture}
\caption{\label{bar_b2} Trajectories of upper and lower skyrmions for b=2 barrier system for various values of $\Gamma$: a)$\Gamma=0.1$, b)$\Gamma=0.25$ , c)$\Gamma=0.3 $.}
\end{figure}

Next we have studied several cases of the scattering of the same two 1-skyrmion configuration off potential barriers. Fig.(\ref{bar_b2}) shows the trajectories of the upper and lower skyrmion of a two 1-skyrmion configuration scattering off a potential barrier of width $b=2$, for various values of $\Gamma$. It can be seen in each plot that the skyrmions are deflected as they traverse the barrier. This deflection always occurs in the direction of the centre of the configuration and hence the normal circular path is deformed as the skyrmions overcome the barrier. Trajectories for a smaller barrier width show a sharper deflection than those with larger $b$. Since the skyrmions are extended objects, when they are traversing the barrier they feel the barrier the most at the middle point. It is then only natural that the maximum point of deviation from the normal circular path will be at this point. Comparing trajectories of the same $\Gamma$ but with different values of $b$, we note that the skyrmions in the system with the larger value of $b$ have more time to adjust to the barrier once they are `on' top of it and therefore their path is not as sharp as for a smaller $b$. The larger value of $b$ `smooths' out the sharpening effects seen in the system with a smaller value of $b$'s. More interestingly, during this deviation the skyrmions speed up as they traverse the barrier. In Figs.(\ref{bar_b2})a and Figs.(\ref{bar_b2})b the times taken for the skyrmion centre to reach the edge of the barrier are 110 secs and 140 secs respectively, but the times taken for the centre to traverse the full width of the barrier are only 25 secs and 12.5 secs.

The deflection and speeding up of the skyrmions can be explained by examining Table(\ref{2S table}) which shows a table of the energies of the two 1-skyrmion configurations for differing values of the skyrmion separation $d$, with the potential coefficient $\gamma_{3}$ set the corresponding value over all the lattice. As explained in section(2), our two 1-skyrmion configuration was constructed in such a way that the skyrmions were in an attractive channel. When they encounter a barrier the energy of their configuration would have to increase. The skyrmions counteract this increase due to the barrier by reducing their separation distance $d$. This is clear by considering the energies of the configuration away from the barrier with $d=6$. The energy of such a configuration is $E=2.1277$. If the same configuration was then placed on the barrier with $\Gamma= 0.2$, the energy becomes $E=2.1283$. Thus any increase in the potential energy due to the barrier, must be compensated by a reduction in $d$. The energy increase of the system would be at its greatest when the skyrmions are in the middle of the barrier hence the biggest deflection is seen at this point. Due to this quick adjustment of $d$, the skyrmions speed up as they traverse the barrier.

\begin{figure}[htbp]
\unitlength1cm \hfil
\begin{picture}(13,13)
 \epsfxsize=10cm \put(0,8){\epsffile{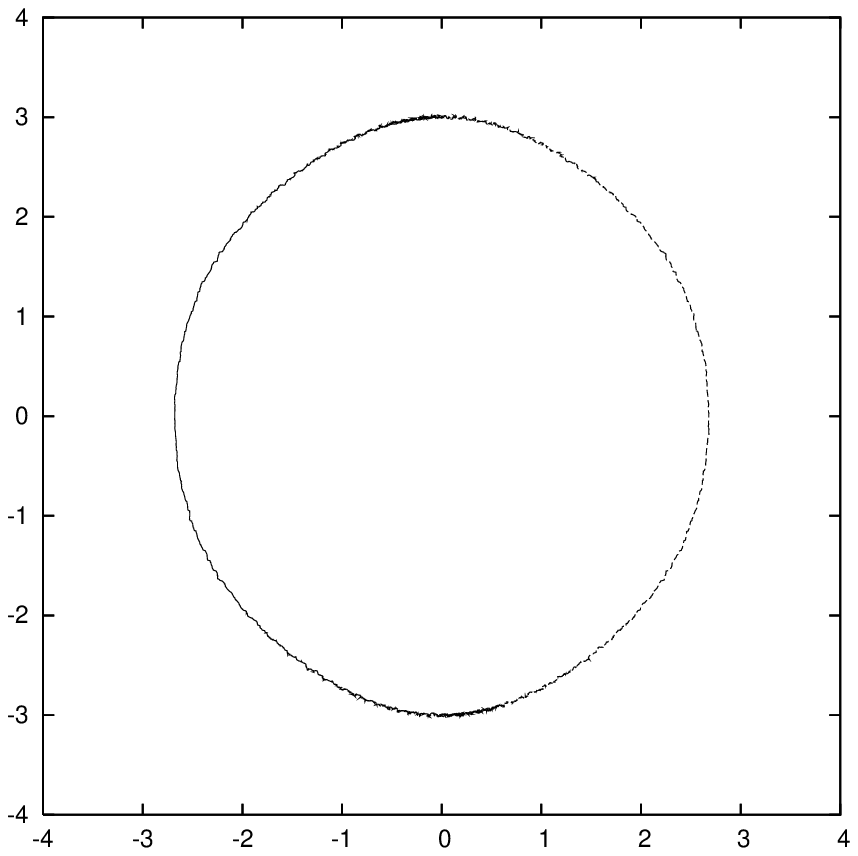}}
 \epsfxsize=10cm \put(0,0.25){\epsffile{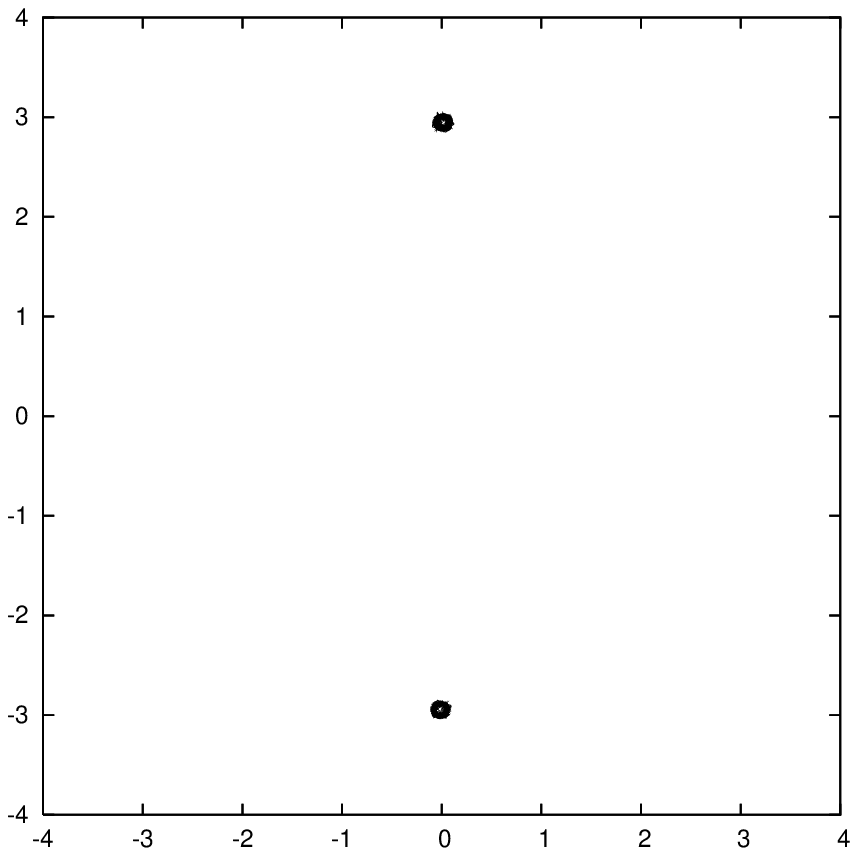}}
 \epsfxsize=10cm \put(7,0.25){\epsffile{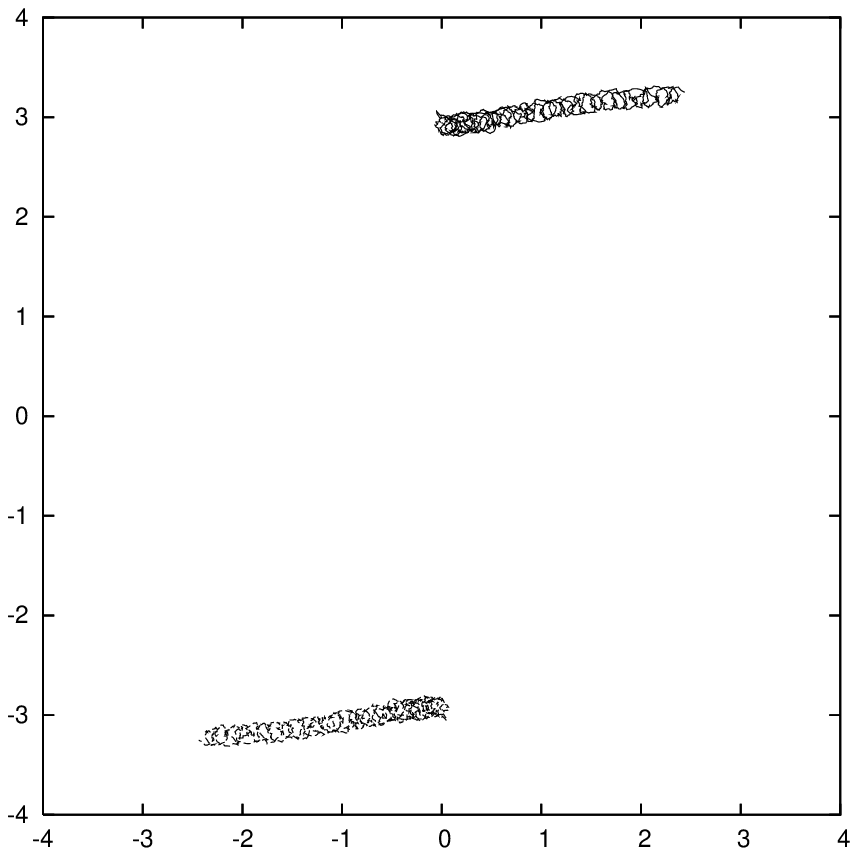}}
\epsfxsize=10cm \put(7,8){\epsffile{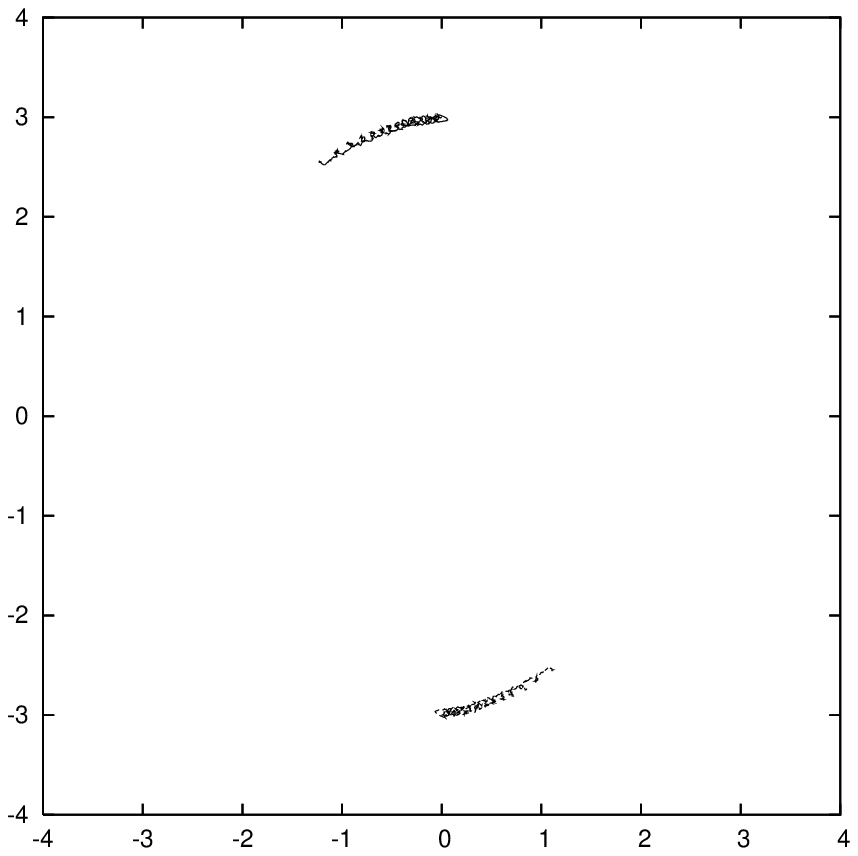}}

\put(3.5,7.5){a}
\put(3.5,0){c}
\put(10.5,0){d}
\put(10.5,7.5){b}
\end{picture}
\caption{\label{Tdynam} Plots showing the transition to asymptotic state b=3 $\Gamma=0.25$,through variations in $\Gamma$ for a fixed b=3: a)$\Gamma=0.10$, b)$\Gamma=0.15$, c)$\Gamma=0.20$, d)$\Gamma=0.25$ .}
\end{figure}

Another interesting feature of the scattering on a barrier is the `transition dynamics' shown in Fig (\ref{Tdynam}). These plots show the transition to a state in which the skyrmions do not traverse the barrier and, instead, move away from each other. The transition to such a state is shown through the variation in the potential coefficient $\Gamma$ for a fixed value of the barrier width $b=3$. Similar plots could also have been obtained by choosing a fixed value of $\Gamma \simeq 0.25$ and increasing the barrier width $b$ from $b=2$ to $b=3$. This effect is due to the binding energies of the skyrmion configuration. Using the previous definition of the binding energy (\ref{EB}) and its constituent parts, one can examine the binding energies in the barrier system. In a system with $b=2$ and $\Gamma=0.1$, $E_{2}=2.1317/8{\pi}$ and $E_{1}=1.0697/8{\pi}$  therefore $E_{B}=-0.0077/8{\pi}$ so the skyrmions are still bound. In $b=2$ and $\Gamma=0.3$, $E_{2}=2.1397/8{\pi}$ and $E_{1}=1.0702/8{\pi}$  therefore $E_{B}=-0.0007/8{\pi}$ and the skyrmions are still bound although a bit more loosely than for the smaller value of $\Gamma$. Next, consider the state where the skyrmions separate from each other i.e. for $b=3$ and $\Gamma= 0.25$. Then $E_{2}=2.1482/8{\pi}$ and $E_{1}=1.0701/8{\pi}$ and therefore, $E_{B}= 0.0079/8{\pi}$. Thus, in this system the skyrmions are no longer bound and separate from each other.


\begin{table}
\begin{center}
\begin{tabular}{|c|c|c|c|c|c|c|}
\hline
$\gamma_{3}$ & d=6 & d=5.5 & d=5 & d=4.5 & d=4 & d=3.5 \\
\hline \hline
1.5 & 2.1279 & 2.1143 & 2.0651 & 1.9795 & 1.9269 & 1.9037 \\
1.4 & 2.1280 & 2.1147 & 2.0659 & 1.9804 & 1.9279 & 1.9046 \\
1.3 & 2.1282 & 2.1150 & 2.0667 & 1.9811 & 1.9288 & 1.9055 \\
1.2 & 2.1283 & 2.1154 & 2.0675 & 1.9824 & 1.9297 & 1.9063 \\
1.1 & 2.1285 & 2.1157 & 2.0683 & 1.9834 & 1.9307 & 1.9072 \\
 1  & 2.1277 & 2.1140 & 2.0643 & 1.9785 & 1.9264 & 1.9028 \\
0.9 & 2.0780 & 2.0628 & 2.0121 & 1.9296 & 1.8807 & 1.8600 \\
0.8 & 2.0283 & 2.0117 & 1.9598 & 1.8808 & 1.8353 & 1.8171 \\
0.7 & 1.9786 & 1.9605 & 1.9075 & 1.8319 & 1.7900 & 1.7743 \\
0.6 & 1.9289 & 1.9094 & 1.8552 & 1.7831 & 1.7447 & 1.7314 \\
0.5 & 1.8792 & 1.8582 & 1.8030 & 1.7342 & 1.6993 & 1.6886 \\

\hline

\end{tabular}

\caption{\label{2S table} Table showing the variation in total energy, in units of $8\pi$, for a two 1-skyrmion configuration with the $\gamma_{3}$ coefficient set at the corresponding value all over the lattice.}
\end{center}
\end{table}

\newpage
\section{Angular momentum }

\begin{figure}[htbp]
\unitlength1cm \hfil
\begin{picture}(6,6)
 \epsfxsize=7cm \put(0,0.5){\epsffile{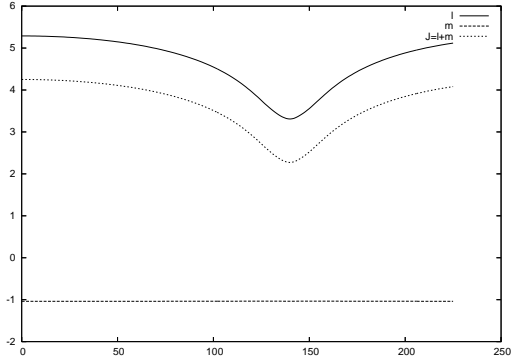}}
\end{picture}
\caption{\label{l b2_025} Plots of the orbital angular momentum $l$, total magnetization $m$ and total angular momentum $J=l+m$ for b=2 $\Gamma=0.25$.
}
\end{figure}

In this section we present the explanation of our results based on the study of the total angular momentum $J$. The orbital angular momentum $l$ and the total magnetization in the third direction $m$ were calculated through all the simulations using the definitions given by (\ref{l}) and (\ref{m}). Fig.(\ref{l b2_025}) shows a plot of $l$, $m$ and $J$ for a two 1-skyrmion configuration interacting with a  potential barrier of width $b=2$ and $\Gamma= 0.25$. It is clear that $\dot{m}=0$ throughout, but that $l$, and $J$, are not conserved in time.



\begin{figure}[htbp]
\unitlength1cm \hfil
\begin{picture}(6,6)
 \epsfxsize=7cm \put(0,0){\epsffile{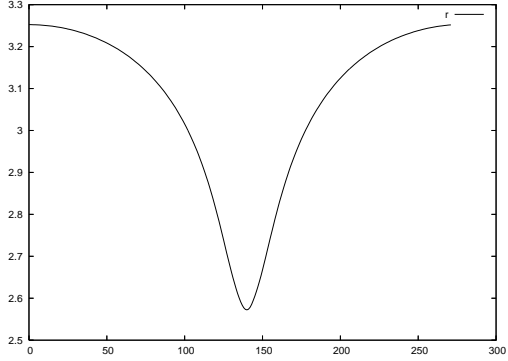}}
\end{picture}
\caption{\label{r b2_025} Plot of the average skyrmion radius as a function of time for a potential barrier with $b= 2$ and $\Gamma = 0.25$.}
\end{figure}

In systems involving a two 1-skyrmion configuration the guiding centre coordinate $\underline{R}$ , defined in (\ref{R}), corresponds to the centre of the configuration. A calculation of $\underline{R}$ during the barrier simulations has indeed shown that $\underline{R}=0$. This is expected since the trajectories of the skyrmions in the system are always symmetric with respect to a reflection through the origin and so the centre of the configuration always lies at this point. Considering (\ref{r1}) with $\underline{R}=0$, we note that the orbital angular momentum $l$ and the average size of the skyrmions $r$ are now directly related to each other in the barrier system by:
\begin{equation}
r^{2}= \frac{l}{2 \pi Q} \label{r} .
\end{equation}
Fig(\ref{r b2_025}) shows a plot of the average skyrmion radius as a function of time for a potential barrier system of width $b=2$ and $\Gamma=0.25$. The points at which $r(t)$ approaches its minimum, corresponds to the skyrmions traversing the barrier. The point at which they have reached the maximum of the barrier corresponds to the minimum of $r(t)$. Thus as the skyrmions traverse the barrier their average size decreases from its starting value by around $\%20$. Since the tail of the skyrmion is exponentially localised and this localisation is governed by the potential coefficient parameter $\gamma_{3}$, it is expected that due to the inhomogenaity in $\gamma_{3}$ their size would decrease in the region of larger $\gamma_{3}$ explaining the observed behaviour in $l$.


\begin{figure}[htbp]
\unitlength1cm \hfil
\begin{picture}(6,6)
 \epsfxsize=7cm \put(0,0){\epsffile{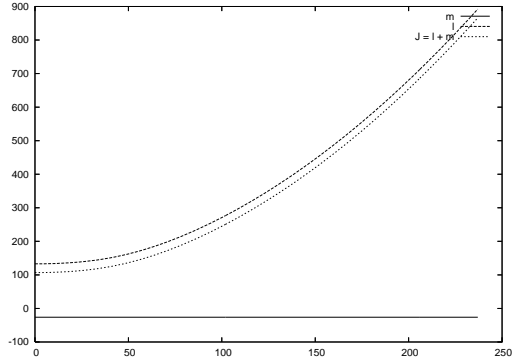}}
\end{picture}
\caption{\label{h b2_025} Plots  of the orbital angular momentum $l$, total magnetization $m$ and total angular momentum $J=l+m$ for a potential hole with b=2 $\Gamma = -0.25$.}
\end{figure}

\begin{figure}[htbp]
\unitlength1cm \hfil
\begin{picture}(6,6)
 \epsfxsize=7cm \put(0,0){\epsffile{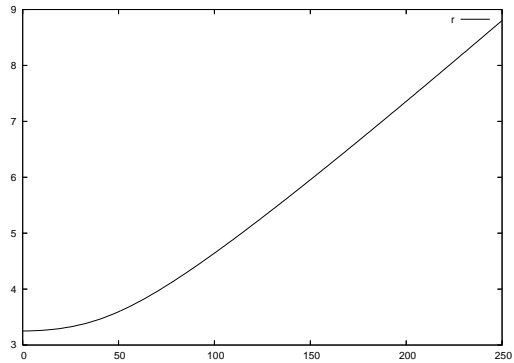}}
\end{picture}
\caption{\label{r h2_025} Plot of the average skyrmion radius as a function of time for a potential hole with $b= 2$ and $\Gamma = -0.25$.}
\end{figure}

Fig(\ref{h b2_025}) shows a plot of $l$, $m$ and $J$ for a two 1-skyrmion configuration interacting with a potential hole of width $b=2$ and $\Gamma= 0.25$. Again, it is clear that $\dot{m}=0$ throughout, but $l$, and $J$, are not conserved in time, analogous to what was seen in the system involving a potential barrier. The guiding centre coordinate $\underline{R}$ for this system can also be shown to vanish and thus (\ref{r}) is valid also in systems with potential holes. Using this, we can therefore plot $r(t)$ for a potential hole. Fig(\ref{r h2_025}) shows a plot of the average skyrmion radius as a function of time for a potential hole system of width $b=2$ and $\Gamma = -0.25$. It is clear from the plot that as the skyrmions approach the boundary asymptotically along the edge of the hole, the average size of the skyrmion increases continually, increasing to approximately 3 times its initial size. This is due to the tail of the skyrmions. The skyrmions cannot penetrate the hole, as explained in the previous sections, but its tail can. The exponential localisation of the skyrmions, as explained earlier, is governed by the potential coefficient $\gamma_{3}$. In the region of reduced $\gamma_{3}$, their average size is able to grow and so it continues to increase until they reach the boundary of the system, where, they get reflected. In our simulations we saw that following this reflection the skyrmions' size decreases back to its starting value as the system tends to its starting point.  


\newpage
\section{Further discussion of Angular Momentum}

Here we discuss further the apparent non-conservation of $J$. Assuming the definitions of $l$, $m$ and  hence of $J$ to be valid and shown to be true in the free system, we ask ourselves can we explain this more qualitatively? Let us consider the behaviour of $J$. The form of $l(t)$ from Fig.(\ref{l b2_025}) indicates that $\dot{l} \ne 0$ and hence $\dot{J} \ne 0$. The calculation of $l$ includes only the contribution of the fields, but clearly this may, for systems involving obstructions, not be sufficient. We can consider adding an external contribution due to the potential obstruction and see whether this restores  $J$-conservation. How the potential obstructions affect, if at all, the orbital angular momentum needs to be considered. The symmetric obstructions can be written as a contribution to the potential term $V(\phi)$ in terms of Heaviside functions. The potential term and this inhomogeneity expressed in terms of Heaviside functions can be written as:
\begin{equation}
V(\phi)=\frac{1}{2}\gamma_{3} \left(1-\phi_{3}^{2} \right) \pm \frac{1}{2}\Gamma \left(1-\phi_{3}^{2} \right)  \left [ \Theta(y+y_{0})-\Theta(y-y_{0}) \right].
\end{equation}
Using the above definition of $V(\phi)$, one can compute the contribution made to $\dot{l}$ in addition to the fields already computed from (\ref{l}). One needs to construct $\dot{q}$ from its constituent parts as shown in (\ref{qdot}):
\begin{equation}
{\dot {q}}=-\epsilon_{ij}\partial_{i}\partial_{l}\sigma_{jl} =-\epsilon_{ij}\partial_{i}\left ( \frac{\delta W}{\delta \underline{ \phi}}\cdot{\partial_{j} \underline \phi}\right ) .\nonumber
\end{equation}
Using the properties of the Heaviside functions and their relations to the $\delta$-function one can show that the potential obstruction's contribution to $\dot{q}$ is given by:
\begin{equation}
\dot{q}=\pm \phi_{3}\partial_{x}\phi_{3}\Gamma \left [ \delta(y+y_{0})-\delta(y-y_{0}))\right ] .\nonumber
\end{equation}
With this expression for $\dot{q}$, we can evaluate the total rate of change of the orbital angular momentum due to the obstruction. We find:
\begin{eqnarray}
\dot{l}&=&\frac{1}{2} \iint_{- \infty}^{\infty}\left( x^{2}+y^{2}\right )\dot {q} \,dx  \,dy  \nonumber\\
&=&\frac{\pm \Gamma}{2}\int_{-\infty}^{\infty} \,dx \int_{-\infty}^{\infty}\left( x^{2}+y^{2}\right ) \partial_{x}(\frac{1}{2}\phi_{3}^{2})\left [ \delta(y+y_{0})-\delta(y-y_{0}))\right ] \,dy  \nonumber \\
&=&\frac{\pm \Gamma}{2}\int_{-\infty}^{\infty} \,dx \left ( x^{2}+y^{2}\right )  \partial_{x}(\frac{1}{2}\phi_{3}^{2}) \Bigg |_{y=y_{0}}^{y=-y_{0}}. \label{ldot_pot}
\end{eqnarray}

\begin{figure}[htbp]
\unitlength1cm \hfil
\begin{picture}(6,6)
 \epsfxsize=7cm \put(0,0.5){\epsffile{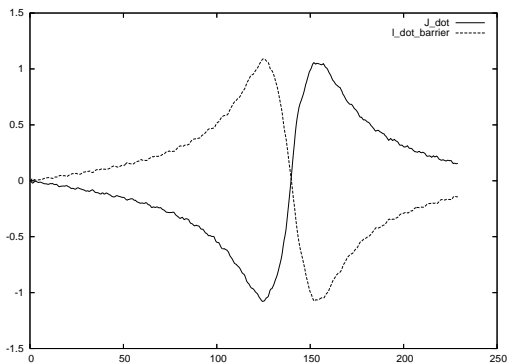}}
\end{picture}
\caption{\label{ldot_barrier} Plots of the numerically calculated time derivative of the total angular momentum $J=l+m$ and the contribution of the barrier to $\dot{l}$ for b=2 $\Gamma=0.25$.}
\end{figure}

The integrals in (\ref{ldot_pot}) have been calculated during each \\ simulation. Fig.(\ref{ldot_barrier}) shows the numerically computed integral contributions due to the obstruction and the numerically calculated derivative of the orbital angular momentum from Fig.(\ref{l b2_025}) plotted with respect to time, for a potential barrier with $b=2$ and $\Gamma =0.25$. It can be seen from the plot that the time evolution of the integral contributions exactly matches that of $\dot{l}$ due to the fields, so that we have:
\begin{eqnarray}
\dot{l}&=&\dot{l}_{fields}+\dot{l}_{barrier}\\
&=&\frac{d}{dt}\Big[\frac{1}{2} \iint_{- \infty}^{\infty}\left( x^{2}+y^{2}\right )q \,dx  \,dy  \Big]+\frac{\pm \Gamma}{2}\int_{-\infty}^{\infty} \,dx \left ( x^{2}+y^{2}\right )  \partial_{x}(\frac{1}{2}\phi_{3}^{2}) \Bigg |_{y=y_{0}}^{y=-y_{0}} \\
&\simeq&0 .
\end{eqnarray}
Due to discretisation effects and numerical inaccuracies, the result is not exact but the qualitative features of the integral contributions makes this a very consistent result.


\begin{figure}[htbp]
\unitlength1cm \hfil
\begin{picture}(6,6)
 \epsfxsize=7cm \put(0,0.5){\epsffile{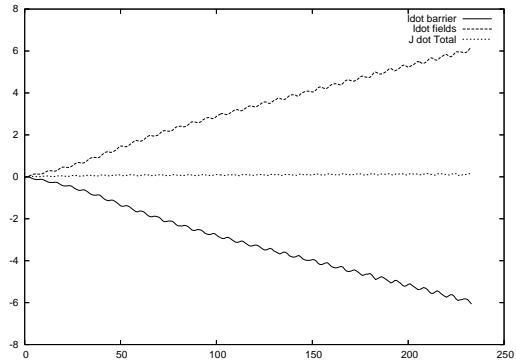}}
\end{picture}
\caption{\label{ldot_hole} Plots of the numerically calculated time derivative of the total angular momentum $J=l+m$ for the fields and the contribution of the hole to $\dot{l}$ for b=2 $\Gamma = -0.25$.}
\end{figure}

 Fig.(\ref{ldot_hole}) shows the numerically computed integral contributions due to the obstruction and the numerically calculated derivative of the orbital angular momentum from Fig.(\ref{h b2_025}) plotted with time, for a potential hole with $b=2$ and $\Gamma =0.25$.
It is clear from these plots that the conservation of the total angular momentum $J$ is restored by the introduction of the terms corresponding to the potential obstructions' contribution to $\dot{l}$.


\newpage

\section{Conclusions}

Our studies have shown that the scattering of baby skyrmions of our model off potential obstructions, for which the dynamics is governed by the Landau-Lifshitz equation, exhibits some nontrivial results.

We have managed to understand quite well the observed scattering properties of our skyrmions despite their, at first sight, somewhat  non-intuitive behaviour. Thus, in the case of a potential hole the skyrmions were unable to penetrate it and so moved parallel to the `x'-axis at a distance $y^{s}_{max}$ from the hole. The energy considerations have shown that the skyrmions, in systems involving a potential hole, were no longer bound and so
could and did move away from each other.
In the barrier systems the skyrmions were able to traverse the barrier. Our simulations have shown that as the skyrmions traversed the barrier their distance of separation $d$ decreased to overcome this increase in potential energy. At the same time the skyrmions sped up as they climbed the barrier. Whether the skyrmions were bound or not could only be determined by the details of the energetics of the system. It was found that the skyrmions were bound for certain low values of the parameters $b$ and $\Gamma$. At higher values they were no longer bound and were free to move away from each other.

 An interesting observation of our simulations was the apparent non-conservation of the total angular momentum $J$ (given its usual definition).
This non-conservation of $J$ was due to the non-conservation of the orbital angular momentum $l$, as we have found that in all of the simulations the total magnetization in the third direction $m$ was well conserved in time. At the same time we showed  that $\dot{l} \ne 0$.
Thinking about this further we showed that when a system possesses potential obstructions these obstructions made a significant contribution to $\dot{l}$.
Hence one has to modify the conventional definition of $l$. We have found this missing contribution and we have shown that
its change compensates $\dot{l}$, resulting in the overall conservation of $l$ and $J$ for the full system. We believe that most of the results presented here form a generic basis for the description of the scattering of baby-skyrmion configurations in Landau-Lifshitz models. This is primarily due to conservation laws of the Landau Lifshitz systems, as constructed by Papanicolaou and Tomaras\cite{papatom}, and the observation that the potential obstructions contribute to the conservation laws of the system.

\end{document}